# How Dark the Sky: The JWST Backgrounds

Jane R. Rigby[1], Paul A. Lightsey[2], Macarena García Marín[3], Charles W. Bowers[1], Erin C. Smith[1], Alistair Glasse[4], Michael W. McElwain[1], George H. Rieke[5], Ranga-Ram Chary[6], Xiang (Cate) Liu[6], Mark Clampin[7], Randy A. Kimble[1], Wayne Kinzel[8], Vicki Laidler[8], Kimberly I. Mehalick[1], Alberto Noriega-Crespo[8], Irene Shivaei[5], Dennis Skelton[9], Christopher Stark[1], Tea Temim[10], Zongying Wei[2], and Chris J. Willott[11]

[1] NASA Goddard Space Flight Center, 8800 Greenbelt Road, Greenbelt, MD 20771, USA; Jane.Rigby@nasa.gov
[2] Ball Aerospace, Retired, 1600 Commerce Street, Boulder, CO 80303, USA
[3] European Space Agency, 3700 San Martin Drive, Baltimore, MD 21218, USA
[4] UK Astronomy Technology Centre, The Royal Observatory, Blackford Hill, Edinburgh, EH9 3HJ, UK
[5] Steward Observatory, The University of Arizona, 933 North Cherry Avenue, Tucson, AZ 85721, USA
[6] IPAC, MC 314-6 1200 East California Boulevard, Pasadena, CA 91125, USA
[7] Astrophysics Division, Mary W. Jackson NASA Headquarters Building, 300 Hidden Figures Way SW, Washington, DC 20546, USA
[8] Space Telescope Science Institute, 3700 San Martin Drive, Baltimore, MD 21218, USA
[9] Sigma Space Corporation, Retired, USA
[10] Department of Astrophysical Sciences, Princeton University 4 Ivy Lane, Princeton, NJ 08540-7219, USA
[11] NRC Herzberg, 5071 West Saanich Road, Victoria, BC V9E 2E7, Canada

Received 2022 November 15; accepted 2023 February 17; published 2023 April 24

## Abstract

We describe the sources of stray light and thermal background that affect JWST observations, report actual backgrounds as measured from commissioning and early-science observations, compare these background levels to prelaunch predictions, estimate the impact of the backgrounds on science performance, and explore how the backgrounds probe the achieved configuration of the deployed observatory. We find that for almost all applications, the observatory is limited by the irreducible astrophysical backgrounds, rather than scattered stray light and thermal self-emission, for all wavelengths $\lambda < 12.5 \; \mu$m, thus meeting the level 1 requirement. This result was not assured given the open architecture and thermal challenges of JWST, and it is the result of meticulous attention to stray light and thermal issues in the design, construction, integration, and test phases. From background considerations alone, JWST will require less integration time in the near-infrared compared to a system that just met the stray-light requirements; as such, JWST will be even more powerful than expected for deep imaging at 1–5 $\mu$m. In the mid-infrared, the measured thermal backgrounds closely match prelaunch predictions. The background near 10 $\mu$m is slightly higher than predicted before launch, but the impact on observations is mitigated by the excellent throughput of MIRI, such that instrument sensitivity will be as good as expected prelaunch. These measured background levels are fully compatible with JWST's science goals and the Cycle 1 science program currently underway.

*Unified Astronomy Thesaurus concepts:* Astronomical instrumentation (799); Sky brightness (1462); Infrared astronomy (786)

## 1. Introduction

How deep JWST can see depends on the darkness of its sky. Emission from our solar system and from the Galaxy can be minimized by pointing at high Ecliptic and high Galactic latitudes. JWST is also susceptible to two important sources of additional background. First, out-of-field astrophysical light is scattered into the beam of the open-architecture telescope, which primarily affects near-infrared ($\lambda < 5 \; \mu$m) wavelengths. Second, in the mid-infrared, JWST detects scattered thermal emission from the sunshield and observatory, as well as thermal self-emission from the telescope mirrors and surrounding frill. The levels of these background components are intimately tied to the science performance.

In the simple case of observations of a point source that is much fainter than the sky background, the integration time required to reach a given signal-to-noise ratio (S/N) will scale linearly with the total background level. Such background-limited observations, as they are termed, comprise a large swath of JWST's science portfolio, including the famous "deep fields." In addition, the mid-infrared background emission has a nonuniform distribution across the detector. This requires either a form of self-calibration (e.g., subtracting dithered images to eliminate the background) or in some cases, obtaining an off-source background measurement, which doubles the total integration time required.

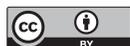

To realize its demanding science goals, JWST had the design objective of being limited, over much of its wavelength range, by astrophysical backgrounds rather than scattering and self-emission. As a result, much care was taken during the design and construction of JWST to minimize the amount of scattered light through extensive baffling, the construction of a large frill around the primary mirror, and scrupulous contamination control, and to minimize the amount of thermal emission through means both passive (the largest example being the sunshield) and active (the MIRI cryocooler).

The deepest near-infrared observations with Hubble (Koekemoer et al. 2013), at 323 ks depth with WFC3-IR in F160W at 1.6 $\mu$m, detected roughly 9000 counts (in e-) for an $m_{AB} = 29.1$ source, and a million counts (in e-) from the background; this high background, via Poisson noise, limits the S/N to ⩽9. By contrast, JWST/NIRCam was expected prelaunch to take only 9.5 ks to reach the same S/N in F150W, with 1600 counts (in e-) from the target and 20,000 counts (in e-) from the background. JWST was expected to be much more powerful than Hubble at these short wavelengths because the aperture is larger. The better-than-expected optical performance of JWST enhances this performance because photometry can be extracted from smaller apertures that have less background. At longer wavelengths, JWST also has a sensitivity advantage over HST/NICMOS due to its much colder temperature. At the longer infrared wavelengths, the previous missions Infrared Space Observatory and Spitzer Space Telescope (Werner et al. 2008) had their sensitivity restricted by confusion noise due to their limited angular resolution; JWST was expected to have immense gains over these missions, because the telescope aperture is much larger. While JWST is warmer than both these missions, for point sources, the higher thermal backgrounds are partially offset by the ability to do photometry in small extraction regions, due to the large telescope aperture. These advantages could be undone if JWST were subject to high levels of stray light or thermal self-emission, for instance, from particulate contamination, inadequate baffling, or thermal shorts.

Since background levels are so closely linked to science performance, characterizing the near-infrared and mid-infrared background levels were key goals of the six-month commissioning period. In this paper, we measure the backgrounds from commissioning observations as well as archival data sets from early Cycle 1 science, and constrain the minimizable backgrounds—the stray light and thermal self-emission.

## 2. The Components of the JWST Background

In this paper, we use the term "backgrounds" in the signal-processing sense: an unwanted excess signal to be removed from the science measurement. Of course, we recognize that geometrically, observatory self-emission always arises in the foreground of the scene, whereas background emission from the solar system and Milky Way galaxy will be in the geometric foreground when the science target is located beyond our Galaxy.

The backgrounds seen in JWST data have multiple origins. Backgrounds are expressed as equivalent units of uniform sky radiance (megaJanskys per steradian, abbreviated MJy/sr) at the JWST focal planes. The expected components of the JWST background, which are plotted in Figure 1, are listed below.

1. In-field zodiacal light: The zodiacal dust within our solar system reflects sunlight at short wavelengths and emits thermally at long wavelengths, with the valley between located at ∼3 $\mu$m. The in-field zodiacal emission depends strongly on ecliptic latitude; it is not expected to vary spatially over the small fields of view of JWST. The in-field zodiacal background varies predictably with date as JWST looks through different path lengths of dust at different temperatures. Figure 1 shows an example of the seasonal variability of the zodiacal background.

2. In-field interstellar medium (ISM): Dust grains within our own Milky Way emit in the JWST bandpass, particularly via prominent emission lines from polycyclic hydrocarbon (PAH) molecules. The in-field Galactic background depends strongly on Galactic latitude, is time invariant, and in practice, has spatial structure over the JWST detector fields of view.

3. Astrophysical stray light: Light from sightlines outside the field of view can enter the optical train by scattering off particulates on the primary or secondary mirrors, or scattering off various observatory structures (Wei & Lightsey 2006; Lightsey & Wei 2012). In general, about half the stray light should arise from sightlines within 35° of the telescope boresight (Lightsey 2016). Figure 2 shows that JWST should be most affected by astrophysical stray light at wavelengths of 1–4 $\mu$m.

4. Observatory stray light: Thermal emission from the sunshield and other parts of the observatory, most significantly, near the deployed tower assembly, can scatter via the primary or secondary mirrors into the field of view. This mechanism is identified as the root cause of the so-called glow stick stray-light features that are seen for MIRI coronagraphy, which are well fit by a graybody spectrum with an effective temperature of $120 \pm 20$ K (Boccaletti et al. 2022), a temperature characteristic of where the deployable tower meets the sunshield. This background has spatial structure across the detector field of view that is independent of where the telescope is pointed.

5. Thermal self-emission from the telescope mirrors: In the mid-infrared, the telescope mirrors and their surrounding frill glow. Since the primary mirror segments have temperatures of 35–55 K, this emission is most dominant for wavelengths $\gtrsim 20\,\mu$m, with the four hottest mirrors, those closest to the sunshield, contributing the most. Before launch, analysis predicted that the mid-infrared background will vary by up to ≈20% for wavelengths $\lambda \geqslant 20\,\mu$m due to changes in the primary mirror temperatures of ∼1 K.

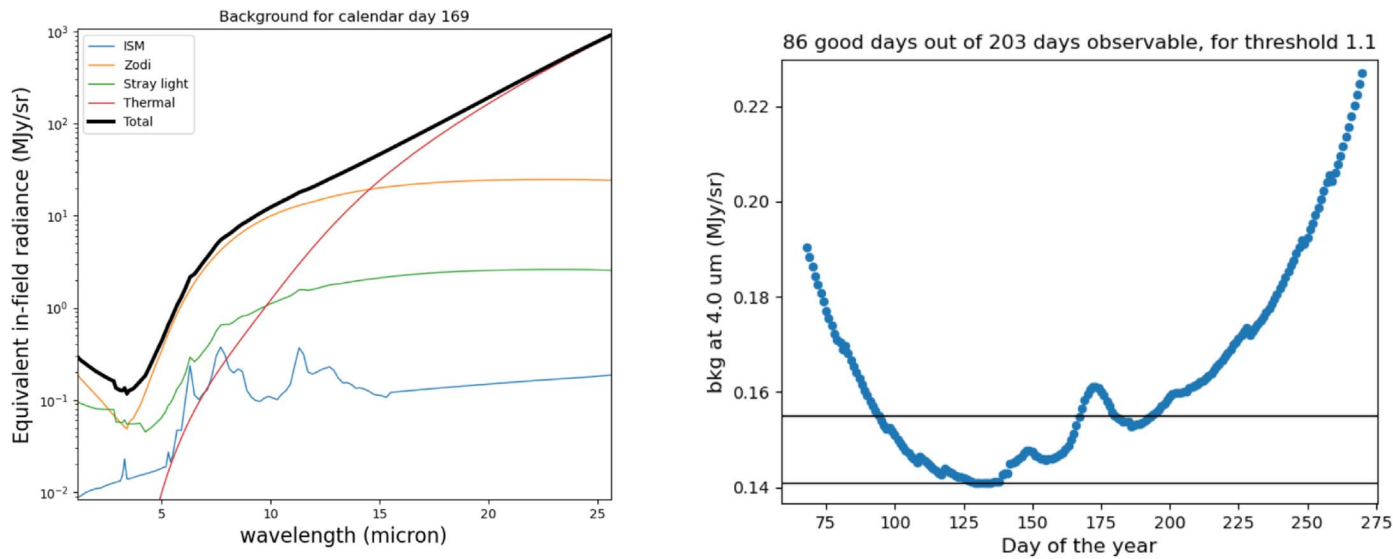

**Figure 1.** The components and variability of the JWST backgrounds. Left panel: predicted components of the backgrounds for the 1.2 minimum zodi benchmark field. The total expected background is shown by the thick black line. From bottom to top at 25 $\mu$m, the subcomponents of the background are in-field interstellar medium (ISM), astrophysical stray light, in-field zodiacal light, and thermal self-emission and scattering. Right panel: predicted variation in the background at 4 $\mu$m for that same benchmark pointing. A horizontal line marks where the backgrounds are 10% above their minimum value. Figure created by the jwst_backgrounds tool for R.A. = 261.68333, decl. = −73.3322, J2000.

The first two of these components were well characterized before launch: we know our own solar system and Galaxy. By contrast, both the amount of astrophysical stray light and the amount of thermal background were quoted before launch as uncertain at the +30% −20% level.[12]

At JWST's longest wavelengths (20–28.5 $\mu$m), the latter two components should dominate compared to the astrophysical backgrounds. Thus, for deep imaging of faint targets, the math is trivial—the integration time required should scale linearly with the thermal background level. If the science objectives for a particular observation require dedicated background frames to remove background spatial structure in the images, then the time required for these frames will need to increase linearly as well.

Shortward of 5 $\mu$m, the prelaunch uncertainty in performance for deep imaging was driven by the uncertain amount of stray light because the other backgrounds were well known. Since the stray light is one of multiple sources of background, a calculation is needed to compute how a change in the stray-light levels affects the total background and thus the integration time. Figure 2 shows this calculation for a benchmark pointing (defined in Section 3.1), in which the relative exposure time required for background-limited observations is plotted as a function of the amount of stray light relative to prelaunch predictions. Were there no stray light added to the natural backgrounds, integration times would be up to 30%–40% shorter; were there twice as much stray light as expected, integration times would be up to 40% longer.

### 3. Design Decisions to Minimize the Backgrounds

Of the many ways in which JWST differs from its predecessors, the most obvious is JWST's open telescope design. Unlike Hubble or Spitzer, JWST does not have a light baffle that encloses the telescope. Instead, JWST uses other structures to control stray light and thermal backgrounds, such as the aft-optics system (AOS), frill, and sunshield (Lightsey & Wei 2012).

Stray-light levels were set that both allowed the required sensitivity to be met and were expected to be technically achievable. To accomplish these stray-light requirements, the general strategy was to restrict the path by which light might arrive at the instrument detectors to the nominal optical path through the telescope only. The AOS, which holds the tertiary mirror and fine-steering mirror (see Figure 2 of McElwain et al., submitted), was covered by a layer of black Kapton. At the front of the AOS, where an intermediate image was formed, a field stop was located to further limit optical access outside the field of view. Each scientific instrument was fully enclosed except for the aperture required to admit light from the telescope. In addition, a pupil stop around the fine-steering mirror eliminates extraneous light from sources outside the telescope light path. Once the four science instruments were mounted into the integrated science instrument module (ISIM), the instrument module was enclosed to exclude celestial light from the vicinity of the instruments. Finally, the sunshield was designed to protect the telescope and instruments from any direct exposure to sunlight throughout the field of regard, allowing them to reach their operating, cryogenic temperatures.

---
[12] https://jwst-docs.stsci.edu/jwst-general-support/jwst-background-model#JWSTBackgroundModel-Uncertaintyinbackgroundlevels, JDox version 1.1

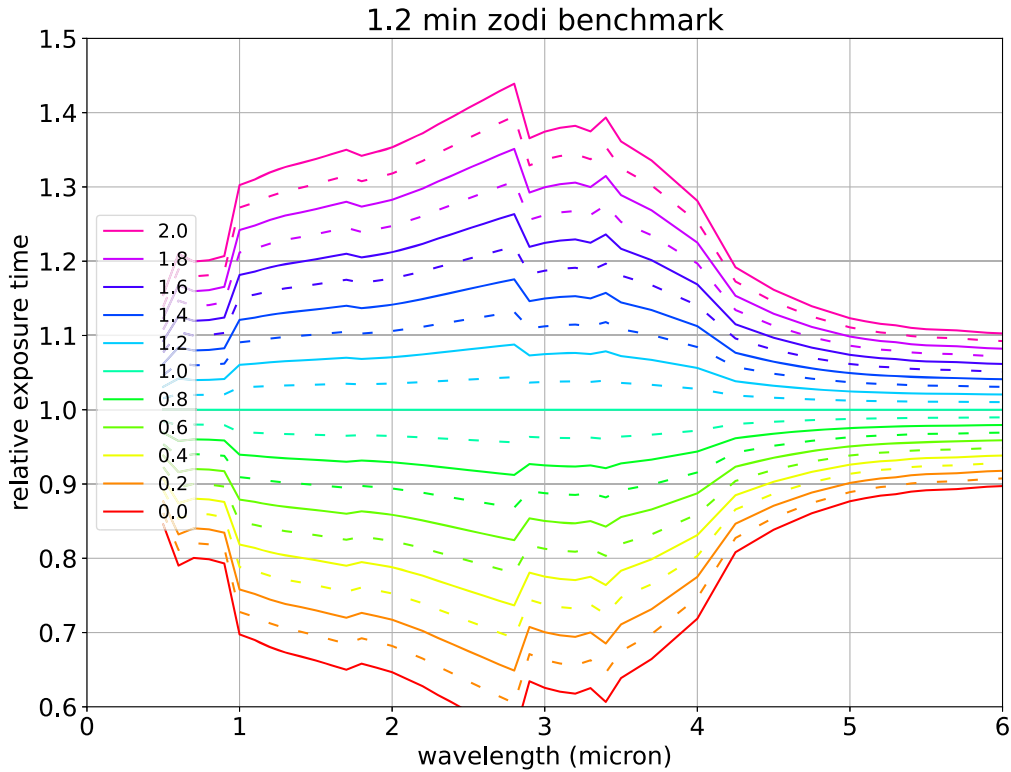

**Figure 2.** The amount of astrophysical stray light affects the integration time required to detect faint targets. Assuming the case of background-limited observations of a point source, the *y*-axis shows how the relative exposure time required to reach a given depth will scale depending on the strength of the stray-light component, which adds onto the natural astrophysical backgrounds, and was the only component at these wavelengths that was uncertain before launch. Curves are relative to the prelaunch prediction amount of stray light for the 1.2 minimum zodi benchmark field—that predicted amount is the flat line at a *y*-value of 1 (green flat curve, marked 1). The bottom (red) curve is for no stray-light component, and the top (fuscia) curve is for twice as much stray light as the prelaunch prediction. Were there no stray light, integrations times could be 60%–70% as long at 1–4 $\mu$m as the prelaunch predictions. Were there stray light at the requirements level (corresponding to the 1.5 × the predicted amount), the integration times would be 17%–20% longer than the prelaunch predictions. Comparing the best-fit stray-light measurement (80% of prelaunch predictions) to the requirements value of 1.5, the integration times realized by JWST in this field at 2.0 and 3.0 $\mu$m should be 78% of what would be the case for a system that just met the stray-light requirements.

### 3.1. Near-IR Contributors and Mitigations

For near-infrared wavelengths (0.6–5 $\mu$m), the astrophysical stray-light requirement was set to be less than the natural background from zodiacal dust. Stray-light modeling, mostly using reverse ray-tracing, with some limited use of forward ray-tracing, was performed by Ball Aerospace and was cross-checked using a different modeling code at NASA's Goddard Space Flight Center. These stray-light models incorporated detailed mechanical models of the structures and material properties of each surface (Lightsey et al. 2014; Lightsey 2016).

To define the stray-light requirement, a benchmark position on the sky was determined that would be a stressing case for stray light; it has a fairly high (−73°) ecliptic latitude, a zodiacal background about 1.2 times higher than the celestial minimum, and (the stressing part) positions the center of the Galaxy toward the front and above the observatory.[13] The celestial coordinates of this 1.2 minimum zodi pointing are given in Table 2. The stray-light requirements were set for this 1.2 minimum zodi field, as listed in Table 1, to be close to but lower than the zodiacal dust contribution at this reference pointing. As discussed below in more detail, the three main contributors to near-infrared stray light are (a) light scattered from particulates on the primary or secondary mirrors, (b) light

---
[13] Galactic longitude and latitude of 319°, −20°; ecliptic longitude and latitude of 262°, −73°.

**Table 1**
Table of Requirements Related to Backgrounds

| Wavelength ($\mu$m) | Requirement (MJy sr$^{-1}$) | Measurement (MJy sr$^{-1}$) | Note |
| --- | --- | --- | --- |
| 2.0 | 0.091 | 0.035 (NIRISS); 0.06 (NIRCam); 0.05 (model scaling) | stray light |
| 3.0 | 0.07 | 0.03 (NIRCam); 0.05 (model scaling) | stray light |
| 10. | 3.9 | range 1–6; best-fit 5 | thermal |
| 20. | 200. | 155 ± 15 | thermal |

**Note.** The stray-light requirements are relative to the benchmark 1.2 minimum zodi pointing. For the stray-light requirements, our measurements are given from the NIRCam and NIRISS instruments, as well as the value that is 80% of the predicted stray-light component of the background model for that pointing on that day of observation. For the mid-infrared, the value quoted is the value of the thermal background spectrum that was fit to the observed data at those wavelengths.

scattering off the secondary mirror support struts, and (c) background from the so-called truant path, defined below.

Particulate scatter, especially off the primary and secondary mirrors, represents a significant fraction of the predicted stray-light background in the near-infrared (see the top of Figure 3). An extensive program of contamination control was instituted to maintain highly clean conditions throughout integration and test, including at the launch site (see Wooldridge et al. 2014a, 2014b). A cover for the entrance to the AOS was only removed when necessary for testing purposes, to keep the tertiary and fine-steering mirrors as well as the science instruments clean. The primary and secondary mirrors were visually inspected for particulates and spot cleaned as needed. This work continued throughout telescope and observatory integration and testing as well as at the launch site. Since in situ contamination measurements were not possible on the mirrors, surrogate contamination wafers were used to monitor the cleanliness of the environment throughout the telescope development. These wafers were placed as close to the flight telescope hardware as possible and used conservative assumptions when applying the measured contamination to the mirrors. For example, wafers were facing up, whereas the telescope optics were often perpendicular to the ground or facing toward the ground to mitigate the contamination accumulation. The stray-light models were used to derive a particulate percent area coverage (PAC) requirement of 1.5% for the primary mirror and 0.5% for the secondary mirror. Contamination monitoring and assumptions for redistribution during launch predicted an end-of-life primary mirror PAC of 0.754% and secondary mirror PAC of 0.149%—both levels are significantly below the requirements.

Molecular contaminants were estimated at 32 Å for the primary mirror and 38 Å for the secondary mirror at launch via ellipsometry on optical witness samples that traveled in close proximity to the flight hardware. Analytical estimates of launch exposure and deposition during cooldown of the observatory predicted totals of 45 Å and 59 Å, respectively, at end-of-life for the sorts of molecules (e.g., organics) commonly referred to as nonvolatile residue (far below the 350 Å requirement). No spectral features have been detected in flight that would indicate unexpected deposition. Similarly, the carefully controlled cooldown of the observatory resulted in predictions of water-ice deposition below the 42 Å requirement on that component for each of these optics. There is no clear detection of spectral features from water ice on the optics in flight, although the breadth of the expected water-ice features and the uncertainties in the intrinsic throughput of the instruments preclude setting precise quantitative limits on the achieved deposition on individual optics. Upper limits on the total equivalent width of the (undetected) strong water-ice absorption feature in the 3.1 $\mu$m region indicate compliance with the total allocation for water-ice deposition on the summed elements of the optical train.

The coverings on the secondary mirror support struts were designed to provide a high degree of thermal stability for the secondary mirror positioned by the long struts. The cover design incorporated a strip of black Kapton on the upper side of each strut, with the rest of the strut profile covered by vapor deposited aluminum (VDA), including particularly the view to the warm sunshield. While meeting the thermal stability requirements, celestial sources can specularly reflect or scatter light from the undersides of the struts onto the primary mirror, which constitutes one of the larger sources of near-infrared background (see Figure 3).

An internal telescope pupil stop was located near the fine-steering mirror. This stop was oversized compared to the image of the primary mirror at the stop to allow the full primary mirror area transmission, and to provide a well-defined pupil for wave-front sensing. However, this left a gap between the stop and the primary mirror pupil image. Light from celestial sources behind the primary mirror that passes near the primary mirror periphery could reflect off the SM and pass through the gap at the stop, creating a stray-light source known as the truant path. A supported, black Kapton frill, extending outward around the periphery of the primary mirror, was installed to substantially block this truant-path light. Approximately 4 m$^2$ of frill area was projected to the internal pupil stop, as compared to 25 m$^2$ for the primary mirror. Particulates on the frill will scatter in the same manner as those on the primary mirror, and frill cleanliness is therefore also important. Two notches in the frill near the upper wing hinges were retained to ensure noninterference during wing deployments. Two other small gaps between the projected frill and the stop remain. These allowed for some margin for sunshield deployment, between the projected sunshield shadow line and the edge of the frill itself. The notches and gaps can be seen as truant-path sources at the bottom of Figure 3. Analysis showed that the residual truant-path contribution, while still one of the larger near-infrared background contributors, would be acceptable (see the bottom of Figure 3). Good alignment between the primary mirror image (and frill) at the fine-steering mirror mask also had to be achieved to provide the maximum suppression of the truant path.

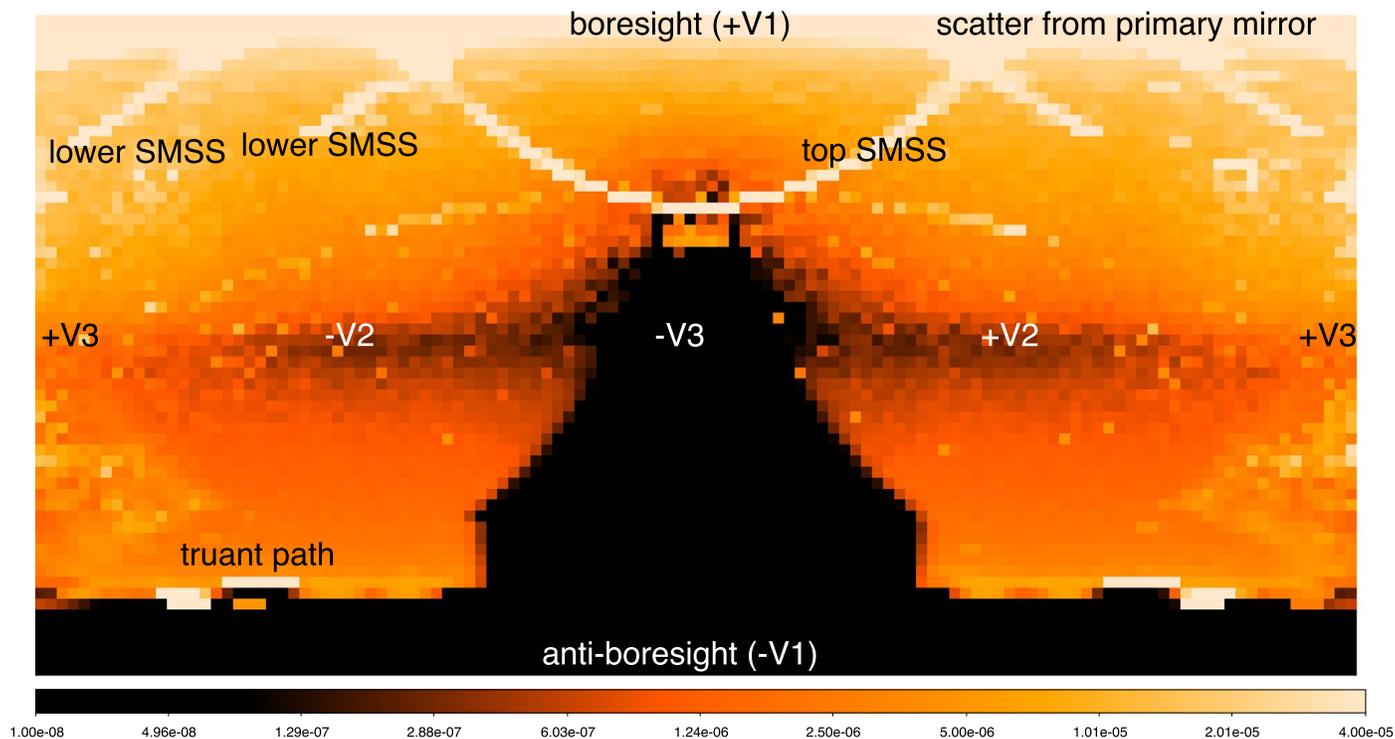

**Figure 3.** Map of the predicted susceptibility of JWST to stray light at 2.0 μm, in spherical coordinates, Mercator projection, and logscale stretch. The map shows the contribution of each direction to stray light. The map is dimensionless because what is plotted is the radiance at the NIRCam detector per unit sky radiance (also known as the radiance transfer function). Stray light is predicted by multiplying this susceptibility map by a map of the sky (for a given pointing, in this coordinate system), and then summing. The vertical direction is the coelevation angle (latitude), from 0° (along the boresight, at the top) to 180° (antiboresight, at the bottom). The horizontal direction is the azimuth angle (longitude), with zero at the −V3 axis in a right-handed sense, and with −180° at left and +180° at right (both are +V3). The bright area near +V1 (coelevation near 0) is due to scattering from the primary mirror and frill. The bright arcs at the top are reflections off the secondary mirror support struts (SMSS). The dark polygonal area is blocked by the sunshield, with very low scattering from the upper sunshield surface. The dark wings extending out from −V3 toward ±V2 show directions parallel to the primary mirror, with a low probability of nearly orthogonal scattering into the line of sight of the telescope. The dark band at the lower edge of the plot (high coelevation, low susceptibility) is due to blocking from the primary mirror and frill in the antiboresight direction, except for a few openings in the frill which allow a very small amount of truant-path background Observatory coordinates are marked (+V1 = boresight, −V1 = antiboresight, −V2 = port, +V2 = starboard, −V3 = "down," +V3 = "up"). The variation with wavelength is small; the maximum value changes by only 11% over 0.7–10 μm. Figure adapted from Figure 3 of Lightsey (2016).

### 3.2. Mid-infrared Contributors and Mitigations

For the mid-infrared, thermal emission from observatory surfaces was expected to establish the background level beyond about 15 μm; shortward of this, the thermal self-emission was expected to be lower than the zodiacal background. Observatory contributions to mid-infrared requirements of 3.9 and 200 MJy sr$^{-1}$ were set at 10 and 20 μm, compared to the expected in-field zodi contributions of 15 and 31 MJy sr$^{-1}$; see Table 1. To predict the mid-infrared background levels, reverse ray-tracing was performed, with structural temperatures provided by thermal models and emissivity and scatter characteristics from material properties.

By design, the large JWST sunshield (see the companion PASP paper by Menzel et al., submitted) keeps the primary and secondary mirrors in the shadow of the sunshield, which allowed them to cool to 35–55 K, establishing the background level for the longest operable wavelength of 28.5 $\mu$m. As expected, of the 18 primary mirror segments, the 4 segments closest to the sunshield have the highest temperatures and therefore contribute most to this long-wavelength background.

The sunshield itself, and particularly the top layer, L5 (the coldest), establishes the background level for shorter (∼15 $\mu$m) mid-infrared wavelengths. Modeling of the deployment established alignment tolerances for the relative positioning of the five layers with respect to each other. This was important to ensure that the sunshield background contribution was only from L5, the coolest layer, and not via scattering from an exposed edge of any warmer layers. In addition, each layer should see only the adjacent layer in order to maintain the appropriate temperature distribution throughout the sunshield assembly; for example, L5 should have a view only to L4, but not to the much warmer L3–L1 layers. A third consideration was maintaining the separation of the sunshield layers to permit sufficient escape of heat to meet the thermal requirements. Design and analysis of the propagation of heat and sunlight along the spreader bars, which cause and maintain the separation of the sunshield layers, ensured that they and their capping epaulets contributed only a low level of stray light.

Thermal emission from the warm region of the deployable tower assembly (DTA) and the core region of the sunshield near the DTA sets the background near 10 $\mu$m. The view from these areas to the secondary mirror is blocked by the bib, a lightweight, black Kapton assembly that is attached to the lower part of the primary mirror backplane and was deployed along with the DTA. The size and final design shape of the bib was a compromise to ensure proper deployment, but still provide a sufficiently small view to the core region to meet the mid-infrared stray-light requirements. Since modeling showed that at equilibrium, the upper portion of this bib would run near 60 K, a smaller, deployable flap was added to block direct visibility to this warm portion of the bib.

The ISIM Electronics Compartment (IEC) was placed within the cold region of the observatory to house critical electronics units operating at 273 K to 313 K. The science instruments had to be isolated from the IEC's thermal emission.

## 4. The JWST Background Model

Figure 1 shows the components of the JWST background model, as well as the expected seasonal variability. Here, we describe how each component is modeled.

1. In-field zodiacal emission—The JWST background model generator (BMG) includes two zodiacal light models, one model from Kelsall et al. (1998), and the other model from Wright (1997) and Gorjian et al. (2000). The latter is what is used operationally. Both models are based on fits to COBE/DIRBE data between 1.2 and 240 $\mu$m, and are extrapolated down to 0.6 $\mu$m.

2. In-field ISM—The JWST background model assumes a fixed ISM spectrum that includes PAH emission, where the intensity is scaled based on the extinction maps of Schlegel et al. (1998), which in turn are based on fits to IRAS and COBE/DIRBE data.

   These first two components of the JWST background model have operational heritage from the Spitzer Space Telescope and Herschel Space Observatory; a standalone version of these first two components is available at https://irsa.ipac.caltech.edu/applications/BackgroundModel/.

3. Astrophysical stray light—This component is predicted by multiplying two monochromatic maps together and summing the result, following Lightsey (2016). The first is a map of the sky in observatory coordinates, centered on the desired telescope pointing with the desired pitch and roll, which includes emission from zodiacal dust and Galactic ISM (each modeled as for the in-field contributions, as well as Galactic stars; model from Wainscoat et al. 1992). The second is a map of the telescope's susceptibility to stray light in observatory coordinates. Susceptibility maps at 3° resolution were generated using a backward ray-trace method (Lightsey 2016). These maps were generated[14] at wavelengths of 0.7, 1, 2, 3, 5, and 10 $\mu$m; the maps were then linearly interpolated on a finer wavelength grid for incorporation into the background model. The 2 $\mu$m susceptibility map is shown in Figure 3.

4. Observatory stray light and thermal self-emission—The observatory integrated thermal model from 2019 was reverse ray-traced to determine the equivalent in-field sky radiance from thermal emission from the sunshield and other parts of the observatory that scatters into the field of view, as well as emission from the telescope mirrors. The hot attitude was assumed (pitch of +5°.2) and the hot season (perihelion). The thermal model includes performance estimates for the beginning-of-life and the end-of-life properties. In the end-of-life case, all of the worst-case conditions were assumed from the thermal liens and threats in order to compare to the requirement value. The reverse ray-tracing predicts the total backgrounds at four discrete wavelengths (10, 15, 20, and 25 $\mu$m), and also summarizes the individual contributions at these four wavelengths from 21 major components of the observatory, such as the secondary mirror, the sunshield, and the deployable tower.

   This reverse ray-tracing process models the observatory at high spatial fidelity, but has a coarse wavelength resolution. Observational planning and comparing to real data requires only coarse resolution as to which parts of the observatory contributed, but requires a spectrum that is finely sampled and comprehensive over the 5–30 $\mu$m range of MIRI. Therefore, for each major component of the observatory, we fit the ray-traced predictions for that component with a blackbody. If the fit converged, we solve for the temperature and effective area times emissivity of that blackbody. Some fits did not converge, which is not surprising becuse each component is a complex system that may not be well modeled by a blackbody. In these cases, we fit a blackbody to each of the six possible wavelength pairs and use the blackbody that fits best, where the goodness-of-fit metric is the fractional error (summing in quadrature) of that blackbody to the predictions at the other two wavelengths. The fits to these components are plotted in panel (B) of Figure 7. This process generated a total predicted background spectrum that closely matched the total predicted backgrounds at the four wavelengths that were ray-traced, and is finely sampled (every 0.1 $\mu$m).

---

[14] Date for reference: 2016 September.

The philosophy of the prelaunch JWST exposure time calculator was to use as-built numbers when they were known, and requirements otherwise. Therefore, conservatively, what was incorporated into the proposal planning system was the thermal spectrum based on end-of-life sunshield performance at a hot attitude.[15] The equivalent spectrum for the beginning-of-life sunshield is used in this paper, but not in the planning system.

The background model is a key part of planning JWST observations. The current version of the background model generator is v4.1. The model is accessed by several tools used by proposers to plan observations:

1. The JWST Backgrounds Tool, or JBT[16], calculates backgrounds for a given sky position. The JBT was used to make Figure 1.
2. The JWST exposure time calculator (ETC),[17] as part of its sensitivity calculations, can include the backgrounds for a specific set of celestial coordinates, either on a given day, or for a typical day.
3. The Astronomer's Proposal Tool (APT)[18] has an option to mark observations as background limited. The JWST planning and scheduling software will schedule these observations days of the year when the time-variable zodiacal background is predicted to be relatively low.[19]

---

[15] It was distributed with the jwst_background tool as filename thermal_curve_jwst_jrigby_v2.2.csv.
[16] https://jwst-docs.stsci.edu/jwst-other-tools/jwst-backgrounds-tool
[17] https://jwst.etc.stsci.edu/
[18] https://www.stsci.edu/scientific-community/software/astronomers-proposal-tool-apt
[19] https://jwst-docs.stsci.edu/jwst-general-support/jwst-background-model/jwst-background-limited-observations

## 5. Methodology

### 5.1. What Can and Cannot Be Measured

Measuring the background levels is one of the simplest measurements that can be made with JWST. In our experimentation, a simple median of imaging data agreed so closely with backgrounds estimated by first removing sources through sigma clipping that we chose to use the simple median as the calculation method.

All we can measure for any pointing is the summed spectrum of all the components listed in Section 2. While what we wish to know are the levels of specific components—the amount of astrophysical stray light, observatory stray light, and thermal self-emission—these photons add to the backgrounds just like the more pedestrian in-field zodiacal emission and ISM light. Therefore, the commissioning observations that were designed to characterize the stray light (PID 01448; PI Smith) observed the total background spectrum for multiple pointings, some that should be dominated by the zodiacal emission, some that should be dominated by Galactic ISM, as well as some relatively deep fields, and the 1.2 minimum zodi benchmark field. The idea was to first characterize that we were correctly measuring the (well-known) zodiacal and ISM emission, and then examine the backgrounds in the deep fields where the stray-light contribution should be a significant fraction of the total backgrounds.

In designing these commissioning activities, we were quite aware that while the telescope is susceptible to astrophysical stray light from many directions (see Figure 3), all that can be measured is the summed stray light from all those directions. The telescope's susceptibility to astrophysical stray light is directional, while the measured backgrounds are integrated over all these directions. We can therefore not hope to empirically create a new stray-light susceptibility map on orbit. Instead, for commissioning, our two goals were to test how well the predicted stray-light model was working, and determine a factor by which to scale the susceptibility map, to better match the on-orbit data, in time to support the Cycle 2 Call for Proposals.

Similarly, while in the models observatory stray light comes from scattering from many different parts of the observatory, all that we can measure is the total mid-infrared spectrum and its time variability. This is why we modeled the observatory stray light as we did, dividing the predicted backgrounds into 21 major components of the observatory—so that we could scale these component spectra up or down to better fit the observed total mid-infrared spectrum. Our goals for commissioning were to determine whether the mid-infrared backgrounds were consistent with prelaunch expectations, identify any exceedances, and determine the on-orbit thermal spectrum in time to support the Cycle 2 Call for Proposals.

In the longer term JWST science data can be harvested to monitor the levels of astrophysical stray light and observatory stray light. This monitoring might hypothetically uncover certain pointings where the astrophysical stray light is

**Table 2**
The Stray Light Fields from the Commissioning Program "Stray Light Pointed Model Correlation," PID 01448 (PI Smith)

| Position | R.A. | Decl. | Date of Observation | Note |
|---|---|---|---|---|
| Galactic Bulge | 17 53 48.0000 | −26 09 0.00 | 2022 May 2 | High Background |
| 1.2 minimum zodi | 17 26 44.0000 | −73 19 56.00 | 2022 May 5 | Benchmark |
| CVZ-South | 06 00 0.0000 | 06 00 0.0000 | 2022 May 5 | Monitoring |
| Low Background 1 | 17 20 0.0000 | +29 59 0.00 | 2022 Apr 29 | Low Background |
| EGS Offset | 14 23 15.0000 | +53 12 30.00 | 2022 May 5 | Low Background |
| Low Background 3 | 08 00 5.0000 | −60 00 10.00 | 2022 May 5 | Low Background |
| Moderate Background | 18 28 0.0000 | −11 00 0.00 | 2022 Apr 29 | |
| High zodi | 08 39 54.7000 | +18 28 36.10 | 2022 Apr 29 | High Zodiacal component |

**Note.** All dates are UTC, and celestial coordinates are J2000.

**Table 3**
The Archival Deep Field Data Sets Used in this Paper

| Deep pointing | R.A. | Decl. | Date of Observation | PID | PI | Category |
|---|---|---|---|---|---|---|
| SMACS J0723.3−7327 | 07 23 18 | −73 27 17 | 2022 Jun 7, 14, 17 | 02736 | Pontoppidan | early-release observations |
| Abell 2744 | 00 14 22 | −30 23 44 | 2022 Jun 28, 29 | 01324 | Treu | early-release science |
| North Ecliptic Pole (NEP-TDF) | 17 22 48 | +65 49 22 | 2022 Aug 26, 30 | 02738 | Windhorst | guaranteed-time observations |

**Note.** All Dates Are UTC, and Celestial Coordinates are J2000.

anomalously high, which might reveal any large deviations from the susceptibility model. Monitoring of observatory stray light might hypothetically reveal any degradation of the sunshield or baffling in the core region.

### 5.2. Sample Selection and Observation Planning

The stray-light commissioning program was designed as eight fields with a range of expected background levels across the sky, observed with shallow integrations of parallel NIRCam and MIRI observations for all fields. Additionally, five of these fields were observed with NIRISS, and three fields were observed with NIRSPEC. Table 2 lists the stray-light commissioning fields and the dates of observation. Observations were taken in MIRI/NIRCAM parallel mode, using three-point MIRI dithers, obtaining eight NIRCam bands (F070W, F115W, F150W, F277W, F356W, F444W, and F480W) with $3 \times 150$ s exposures in SHALLOW 4 mode, and also obtaining four MIRI bands (F560W, F770W, F1000W, and F1280W) with $3 \times 139$ s exposures in FASTR1 mode. Parallel mode resulted in a small offset between the MIRI and NIRCAM fields. As the fields were selected for background measurements (rather than specific features of an astronomical field), these offsets did not affect the data analysis or results. The NIRISS observations were single undithered 330 s exposures in NISRAPID readout mode for each of the F115W and F200W filters.

To supplement these dedicated stray-light observations, we added archival data covering three deep fields. We choose three data sets that should have low backgrounds, which were observed early in Cycle 1, and that have several-hour integrations in several NIRCam and NIRISS broadband filters, and where the data are already public. These deep fields are listed in Table 3.

The "MIRI Sky Background and Variation with Telescope Pointing" commissioning program measured the background seen by the MIRI imager through filters spanning the 5–28 $\mu$m wavelength range at the nominal hot and cold observatory pointing attitudes listed in Table 4. These observations were conducted as part of the commissioning test of thermal stability, in which the observatory was allowed to thermally equilibrate at a relatively hot attitude for 4 d, and then slewed to a cold attitude and again allowed to equilibrate for 7 d (see Section 5.4.4 of McElwain et al., submitted.) Data from a blank region of the sky in both attitudes were obtained using the F770W, F1280W, F1500W, F1800W, F2100W, and F2550W MIRI filters, with exposures of about 850 s in FASTR1 and a three-point dither using the cycling pattern. A $2 \times 2$ mosaic with 50% overlap in the F1800W filter was obtained to evaluate the presence of spatial structure in the thermal emission at longer wavelengths, and to correlate it with the measurements in all other filters. Table 4 gives pointing information and the dates of the observations.

Table 4
Pointings in the "MIRI Sky Background and Variation with Telescope Pointing" Commissioning Program, PID 01052 (PI Glasse)

| Position | Pitch | R.A. | Decl. | Date of Observation |
|---|---|---|---|---|
| Hot Attitude | −0.1 to +0.5 | 18 05 59.9000 | +61 00 29.40 | 2022 May 11 |
| Cold Attitude | −41 to −40.5 | 16 20 10.3800 | +29 31 51.20 | 2022 May 13 |

**Note.** Dates are UTC, celestial coordinates are J2000, and pitches are in degrees, with positive pitch toward the Sun.

To examine the variability of the long-wavelength background at 25.5 $\mu$m, we collect background measurements from ten commissioning, calibration, and Cycle 1 programs, as listed in Table 5.

### 5.3. Photometry

We measured the JWST backgrounds from commissioning and early Cycle 1 data as follows.

For both NIRCam and NIRISS imaging data, we downloaded Level 1b data (uncalibrated files, labeled "UNCAL") from the JWST archive, and reduced the data using the first stage of the JWST pipeline with stock parameter files to create Level 2a (count-rate files, labeled "RATE") files in units of digital numbers per second (DN/s). The JWST calibration pipeline version was 1.7.2. The CRDS context using the PUB server was jwst_0967.pmap.[20]

To convert from measured count-rates (DN/s) into the flux-calibrated units of MJy/sr (a flux density surface brightness), we followed one path for NIRCam and a different path for NIRISS. The reason for this was that the JWST initial flux calibration of NIRCam was unsettled for much of this analysis; it converged in 2022 October. For NIRISS imaging data, we used stage 2 of the JWST pipeline (same pipeline version and CRDS server as above) with the stock parameter files to generate Level 2b files (calibrated files, labeled "CAL") in units of MJy/sr. We performed all photometry of NIRISS images on these flux-calibrated data products. We verified with the NIRISS instrument development team that this approach uses the latest available flux calibration. For NIRCam imaging data, we performed all photometry on the count-rate images (Level 2a), and then converted the measured count-rates into DN/s to surface brightness in MJy/sr by using the PHOTMJSR flux calibration conversion factors that were delivered to CRDS on 2022 October 3,[21] some of which are described in Boyer et al. (2022). To emphasize, we did not use the PHOTMJSR values from the headers because they had not yet been updated at the time this analysis was performed.

Table 5
Commissioning, Cycle 1 Calibration and Cycle 1 Programs Whose 25.5 $\mu$m Data Have Been Used for the Sole Purpose of Characterizing Mid-infrared Background Levels and Variability

| PID | PI | MIRI Filter | Start Date of Observations |
|---|---|---|---|
| 1024 | Ö. Hunor-Detre | F2550W | 2022 May 5 |
| 1027 | M. García Marín | F2550W | 2022 May 25 |
| 1028 | P. Guillard | F2550W | 2022 May 25 |
| 1039 | D. Dicken | F2550W | 2022 May 7 |
| 1040 | Ö. Hunor-Detre | F2550W | 2022 May 25 |
| 1051 | Ö. Hunor-Detre | F2550W | 2022 May 31 |
| 1052 | A. Glasse | F2550W | 2022 May 11 |
| 1521 | M. García Marín | F2550W | 2022 Sep 6 |
| 1947 | D. Milisajvlevic | F2550W | 2022 Aug 2 |
| 2666 | O. Fox | F2550W | 2022 Sep 20 |

Similarly, for MIRI imaging, the Level 1b data, uncalibrated files, in units of DN were processed to Level 2a files (rate files in DN/s) and finally to 2b (CAL) flux-calibrated data, in units of MJy/sr, using the JWST calibration pipeline version, and the 1.6.2 CRDS context using the PUB server was jwst_0969.pmap.[22]

For imaging data, for each exposure and each detector, we took the median value of all illuminated pixels as the background. We then took the median value of these measurements over the multiple exposures per detector, with the standard deviation as the plotted error bar. In the case of NIRCam short wavelength, which has eight detectors, we label the median for each detector (labeled as module A or B, and numbers 1–4), and also plot the median over all detectors (plotted as purple stars.)

The JWST flux calibration at the time of this analysis was based on preliminary in-flight calibration data, not ground-test data. The NIRCam calibration used in this work was based on only two stars (Boyer et al. 2022). The MIRI calibration used in this work was based on 1–2 stars depending on the filter. This is compared to the JWST Cycle 1 calibration plan, which calls for at least nine stars (generally three from each of these categories: A dwarfs, hot stars, and solar analogs) to be observed by the end of cycle 1 (Gordon et al. 2022). For the calibration context used in this work, the flux calibration uncertainties are thought to be at the level of ∼5% for MIRI

---

[20] https://jwst-crds-pub.stsci.edu/context_table/jwst_0967.pmap. The PUB CRDS server has since been deprecated; we believe the equivalent CRDS context on the OPS server is jwst_0988.pmap.
[21] https://jwst-crds.stsci.edu/context_table/jwst_0989.pmap on the OPS CRDS server.

[22] The PUB CRDS server has since been deprecated; we believe this context is equivalent to jwst_0990.pmap on the OPS server.

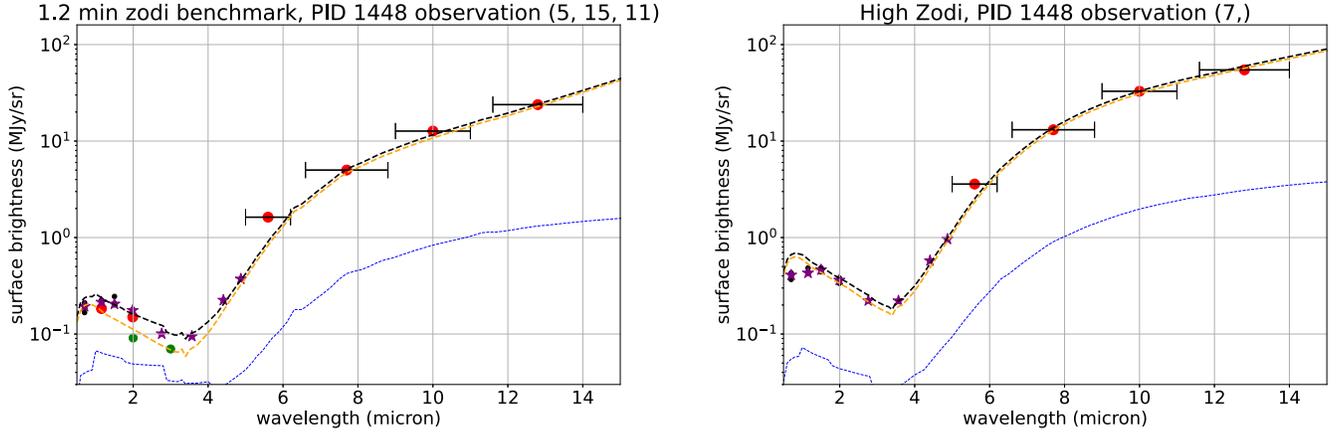

**Figure 4.** The measured backgrounds for two JWST commissioning fields from 0.7 to 13 $\mu$m in a low-background field and a high-background field. On the left are results for the 1.2 minimum zodi field, which was the benchmark field for astrophysical stray light. On the right is the same for the High zodi pointing. Plotted are the measured backgrounds in broadband imaging for three science instruments: NIRISS (red circles <2 $\mu$m), NIRCam (purple stars), and MIRI (red circles >5 $\mu$m). The approximate widths of the MIRI bandpasses are marked with the horizontal error bars. The expected backgrounds from the prelaunch background model are shown by the dashed lines: black for the total background, yellow for the in-field background, and blue for the stray-light component. In the left panel, green circles mark the requirements values for stray light; the amount of predicted stray light (blue curve) is well below the requirements. These plots show that the observed backgrounds agree very closely with prelaunch predictions.

(K. Gordon 2023, private communication) and no more than 5% for NIRCam (M. Boyer 2023, private communication).

### 5.4. Predicted Backgrounds

We predict the JWST background levels using the JWST_backgrounds tool; the inputs are the celestial coordinates and date of observation. For cases where observations were spread over multiple days, we plot the predicted backgrounds for each day the target was observed in the figures.

### 6. Results

Figure 4 shows the measured backgrounds for two fields from the stray-light commissioning program. Results for all the stray-light commissioning pointings are plotted in the Appendix in Figure 9. At a high level, these figures illustrate that the JWST backgrounds overall agree closely with prelaunch predictions. This was not at all a foregone conclusion, as discussed in Section 3, given the prelaunch uncertainty in the levels of stray light and thermal background. We now compare the observed results in detail to the prelaunch expectations.

### 6.1. Astrophysical Stray Light from the Commissioning Fields

Five of the commissioning stray-light pointings were predicted to have low background levels (below 0.3 MJy sr$^{-1}$ at 1–5 $\mu$m). All five of these pointings tell a consistent story: at almost all wavelengths from 1–4.5 $\mu$m, the observed backgrounds are lower than expected. Further, the overlapping photometry from NIRCam and NIRISS at $\lambda = 1.1$ and 2.0 $\mu$m agrees fairly well; this is encouraging because their flux calibrations were determined largely independently.

There is no evidence that the predicted in-field backgrounds (yellow curves in Figure 4) are incorrect, except at the shortest wavelengths ($\lambda < 1$ $\mu$m). Indeed, Figure 4 shows that the NIRCam photometry closely matches the expected shape and normalization of the zodiacal emission at 3–5 $\mu$m, including the valley between the emission from reflected sunlight and the emission from sunlight that was thermally reprocessed by dust.

The simplest explanation for why the 1–3 $\mu$m points in Figure 9 lie below the prediction, is that there is less astrophysical stray light than expected. Scaling down the predicted stray-light component by 20% provides a better fit to the commissioning stray-light data. We will explore this scaling factor in more detail in the next subsections.

The shallow observations of the eight commissioning stray-light fields also show that the total background scales as expected with ecliptic latitude and Galactic latitude.

At the shortest wavelengths ($\lambda < 1.2$ $\mu$m), there appears to be an issue with the background model: the shape of the zodiacal spectrum may be incorrect. This is most obvious for Abell 2744 (Figure 6 panel (A)), where the observed data are below the predictions from zodiacal emission (yellow curve); it also is obvious in the shortest wavelength data points of several of the commissioning stray-light fields (Figure 9). The model at these wavelengths was extrapolated from COBE/DIRBE data at 1.2 $\mu$m; it appears there may be an issue with the scattering function that was assumed in that extrapolation. We plan to investigate this issue and address it in the background model to support future proposal calls.

A second issue affecting short wavelengths ($\lambda \leqslant 1.5$ $\mu$m) is that the observed backgrounds in the Galactic bulge field are much lower than the predictions. We suspect that this arises because JWST resolves much of the faint background into individual stars.

### 6.2. Astrophysical Stray Light in the Benchmark Field

Since the stray-light requirements were designed around a particularly stressing pointing, the 1.2 minimum zodi benchmark, we devote specific attention to the results in that field. Figure 5 compares the predicted amount of stray light for the 1.2 minimum zodi benchmark field (blue line) to the maximum amount of stray light that would have met requirements (green points at 2.0 and 3.0 $\mu$m). Subtracting the expected zodiacal and Galactic contributions, we infer for this field at 2.0 $\mu$m a stray-light level of 0.06 MJy sr$^{-1}$ from NIRCam and 0.035 MJy sr$^{-1}$ from NIRISS. Scaling the model-predicted value by 80% yields 0.048 MJy sr$^{-1}$, which splits the difference between the measured values from NIRCam and NIRISS. Similarly, at 3 $\mu$m, we infer a stray-light level of 0.03 from NIRCam compared to 0.051 MJy sr$^{-1}$ by scaling the stray-light prediction by 80%. These values are listed in Table 1. All these values are lower than the requirements.

The 1.2 minimum zodi field was observed on day-of-year 123, a day when the backgrounds were predicted to be relatively low. According to the background model, had the field been observed on the day with the highest predicted background, the stray-light levels would have been 36% higher at 2 $\mu$m. This is still lower than stray-light requirements, which are 48% higher at 2 $\mu$m than what was predicted for the day of observation. Thus, given the observed measured amount of stray light on the day the field was observed, the stray-light model predicts that there is no observable day when the stray light would exceed the requirements in the benchmark field.

### 6.3. Astrophysical Stray Light from Deep Fields

The stray-light commissioning observations were intended to test whether the stray light behaved as expected with ecliptic latitude and Galactic latitude, and whether the stray-light levels met requirements. As the previous section makes clear, the answers are Yes in both cases. Given this experimental design, these shallow (3 × 150 s) NIRCam observations are not ideal to precisely measure the amount of stray light because they suffer from significant 1/$f$ noise. Therefore, we now examine the near-infrared backgrounds in three deep pointings from early in Cycle 1, with integration times of hours rather than minutes.

Figure 6 shows the measured backgrounds in the three deep fields, sorted from highest to lowest background level. These observations were not designed to measure backgrounds, but they certainly serve the purpose. The results are consistent with what was seen in the commissioning fields, with greater precision due to the long integrations: the backgrounds observed in almost all wavelengths are lower than predicted; NIRISS and NIRCam photometry agree extremely well for wavelengths of overlap; and at 4.1 and 4.4 $\mu$m where the zodiacal emission should be much brighter than any astrophysical stray-light contribution, the photometry is indeed quite close to (but generally lower than) the total predictions, which is exactly what one would expect if the photometry is accurate and the stray-light levels are lower than expected.

Indeed, the North Ecliptic Pole pointing was chosen because it is one of the darkest parts of the infrared sky (Windhorst et al. 2022). As such, panel (C) of Figure 6 indicates the blackest the sky can appear to JWST.

While two of the three deep fields are centered on lensing clusters, with intracluster light that could potentially bias the backgrounds high, the similar backgrounds measured from the two NIRCam modules (one centered on the cluster in each case) indicates that the intracluster light does not significantly bias the measurement of the median background.

We scale the stray-light component of the backgrounds for each field to better match the photometry. The rough scaling factors are 0.4 for Abell 2744, 0.8 for SMACS0723, and 0.9 for NEP-TDF. These scalings are consistent with the 0.8 scaling factor found for the commissioning fields. The result is clear: over the full range $0.7 \leqslant \lambda \leqslant 4.6\ \mu$m, astrophysical stray light has been successfully controlled such that it is less important than the unavoidable in-field astrophysical backgrounds. In other words, as it was designed to be, JWST is limited by the zodiacal light, rather than by stray light, for $\lambda < 5\ \mu$m.

### 6.4. The Mid-IR Backgrounds

We now consider JWST backgrounds for $\lambda > 5\ \mu$m. Figure 7, panel (A), shows the results of the thermal stability test. The photometry is almost identical for the hot and cold pointings.

The backgrounds of long-wavelength MIRI images show low-frequency spatial structure at the 2% level that is independent of the pointing direction. In the future, this structure may be removed by subtracting a noiseless model of the structured background. Depending on the science goals, including the target's spatial extent, Cycle 1 observers should consider choosing a dither pattern or obtaining off-target background images at the same exposure time to correct the science images by subtraction. The total observing time will increase depending on the strategy employed. This spatial structure is a complexity of the backgrounds realized on-orbit; prelaunch analysis assumed the stray-light illumination would be uniform.

In panels (B) and (C) of Figure 7, we subtract the expected astrophysical components of the background from the measured photometry to estimate the residual thermal component. Overall, the observed thermal background spectrum closely matches the prelaunch predictions (blue and red curves in panel (C) of Figure 7), which is testament to the thermal engineering of JWST, in particular, the integrated thermal and optical modeling of a complex system.

There may be some excess emission at 6–15 $\mu$m compared to prelaunch predictions (see panel (B) of Figure 7). The thermal model is made of 16 nonzero subcomponents (colored thin curves) that we derived from the reverse ray-traced models. The observational constraints available are not nearly sufficient to rigorously fit the photometry with this model. Instead, we adjust the prelaunch end-of-life model in an ad hoc manner by adjusting the temperatures and effective areas of three of these components (the sunshield, the primary mirror, and the DTA tower) to better fit the observed residual backgrounds. This ad hoc revised thermal model is plotted as the black curve in panel (C) of Figure 7. This revised thermal curve currently is the best characterization of the JWST thermal spectrum for observation planning and scheduling.

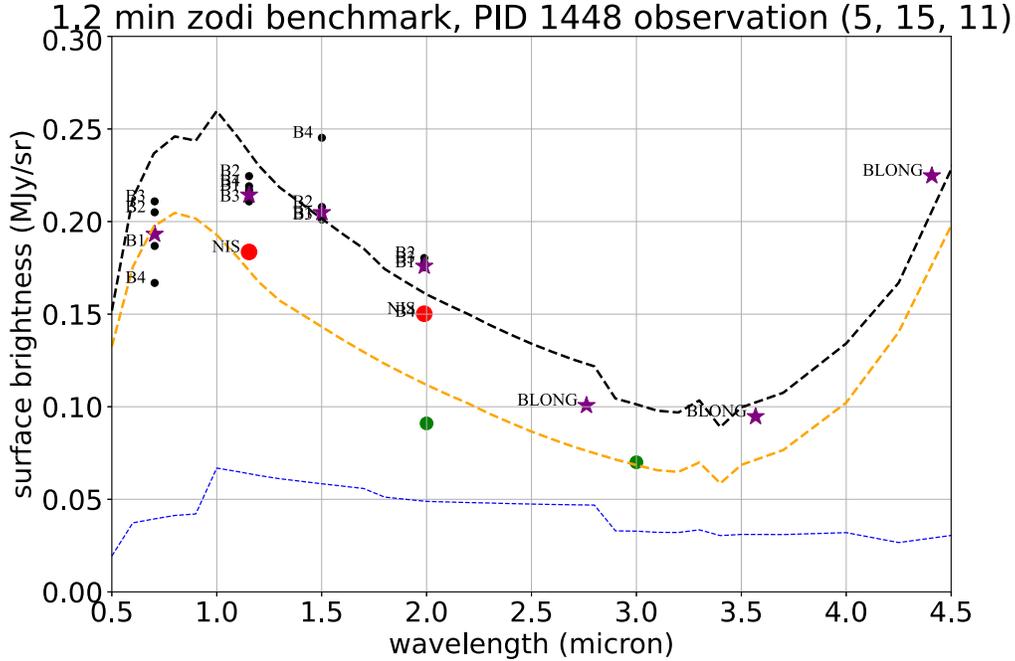

**Figure 5.** Backgrounds for the 1.2 minimum zodi benchmark pointing from commissioning observations. Symbols are the same as in Figure 4. Each detector is labeled (B1–B4 for the NIRCam short-wavelength detectors in module B; BLONG for NIRCam long-wavelength detector in module B, and NIS for the NIRISS detector). The chip-to-chip offsets are much larger than the statistical measurement uncertainties, indicating the importance of remaining systematic issues with calibration.

JWST carried a level 1 requirement to provide the thermal environment needed such that the imaging science instruments would be limited by zodiacal light background over the wavelength range 1.7–10 $\mu$m. The green curve in panel (C) of Figure 7 shows that this is indeed the case for $\lambda \leqslant 12.5\ \mu$m.

JWST carried level 2 requirements, listed in Table 1, for the mid-infrared thermal emission at 10 $\mu$m and 20 $\mu$m. Given current constraints on the thermal spectrum, we estimate the 10 $\mu$m thermal background as in the range 1–6 MJy sr$^{-1}$. With the current amount of precision available from subtraction of the astrophysical backgrounds, and the current uncertainty in the MIRI imaging flux calibrations, it is not yet clear whether the requirement is met at 10 $\mu$m; a larger number of archival data sets from Cycle 1 will need to be analyzed. At 20 $\mu$m, the requirement is clearly met at present, with an estimated equivalent monochromatic flux density of $155 \pm 15$ MJy sr$^{-1}$.

In the case of background-limited observations, integration time will scale with total background level. Panel (D) of Figure 7 shows the expected change in background levels and therefore exposure times due to predicted degradation of the sunshield over time (red curve), as well as the difference in exposure times when using the thermal spectrum that was adjusted to fit the commissioning data (black curve), rather than the thermal spectrum that was assumed prelaunch. This plot shows that for these two beginning-of-life and end-of-life models, predicted degradation of the sunshield over the mission lifetime should increase exposures times by up to 25%. This scaling would place the 20 $\mu$m measured background just below the requirement value at the end of life.

We also infer that, considering backgrounds alone, the exposure times for background-limited observations at $\lambda \sim 10$ $\mu$m may need to be 30% longer than prelaunch expectations, due to an apparent excess of short-wavelength emission. However, the better-than-expected overall throughput of MIRI (see Wright et al. 2023) may largely compensate for this effect.

These results are based on the latest calibrations and reference files available at the time of submitting this paper. We expect the calibration files to evolve and converge as Cycle 1 progresses, and therefore, small variations from the reported mid-infrared background measurements are expected.

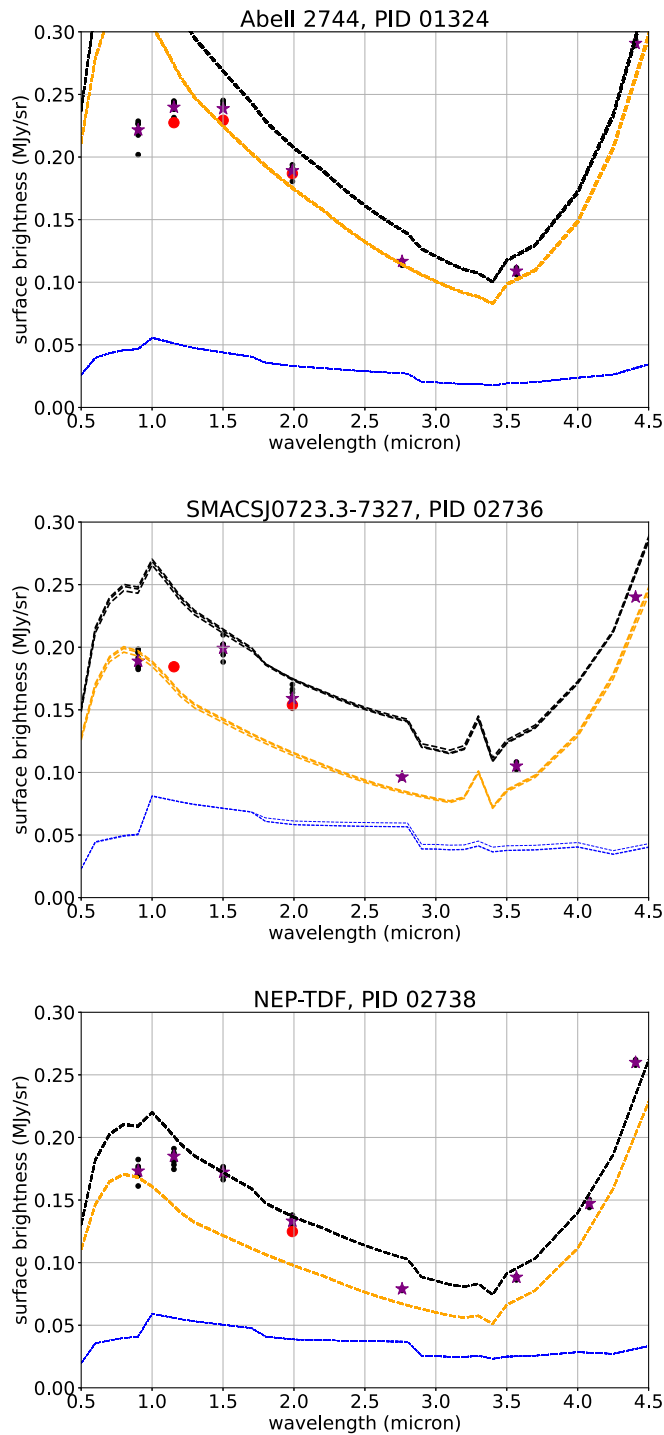

**Figure 6.** The measured backgrounds for the deep observations. Symbols and curves are the same as in Figure 4. Since the zodiacal background is time variable (and therefore the stray light is as well), the predicted background is plotted for each day the field was observed.

*6.5. Variability of the Mid-infrared Background*

The amount of solar energy input to the hot side of the sunshield must vary as the cosine of the pitch angle; it is not obvious how this variation in input energy translates into the temperatures of the cold side of the sunshield and the primary mirrors, both of which should dominate the mid-infrared backgrounds.

Figure 8 shows that the measured background variation in the F2550W filter is about 7%. We searched for but did not find a correlation between this background and the telescope orientation relative to the Sun. Thus, we conclude that the mid-infrared backgrounds are quite stable with time; they still vary enough that the variation must be accounted for in the data reduction. The MIRI team will trend the long-wavelength backgrounds throughout the life of the mission.

## 7. Discussion and Conclusions

Key to the science performance of JWST was the simple question: How dark are the darkest places on the sky to JWST, and in consequence, how quickly will JWST integrate? The main conclusions of this paper are listed below.

1. The level of astrophysical stray light affecting JWST has been measured; it is substantially lower (better) than requirements, and approximately 20% lower than pre-launch predictions.
2. JWST's thermal background spectrum has been well measured; the background level at long ($\sim 20\,\mu m$) wavelengths closely matches prelaunch expectations for the end-of-life performance; there appears to be modest excess at shorter wavelengths (5–15 $\mu m$), whose origin is not yet understood.
3. The variability of the long-wavelength background is 7% at 25 $\mu m$; it is more stable than prelaunch expectations, which may be due to the stability of the primary mirror temperatures, which dominate the thermal emission at the longest wavelengths.
4. Given these measurements, we conclude that JWST is limited by zodiacal emission, not by stray light or its own self-emission, for all wavelengths $\lambda < 12.5\,\mu m$. The one exception is MIRI coronagraphy, where a glow stick stray-light feature is effectively mitigated by the addition of dedicated background observations. Thus, JWST meets its level 1 requirement for backgrounds.

The result that the 25 $\mu m$ background is time variable at only the 7% level, with no obvious dependence on Sun angle, is at first glance surprising. However, this result is also consistent with the small temperature changes seen during the thermal stability test, where the mirrors only changed temperature by 0.2 K between the hot and the cold attitude. Clearly, the mirror temperatures and the mid-infrared backgrounds are more stable with time than expected.

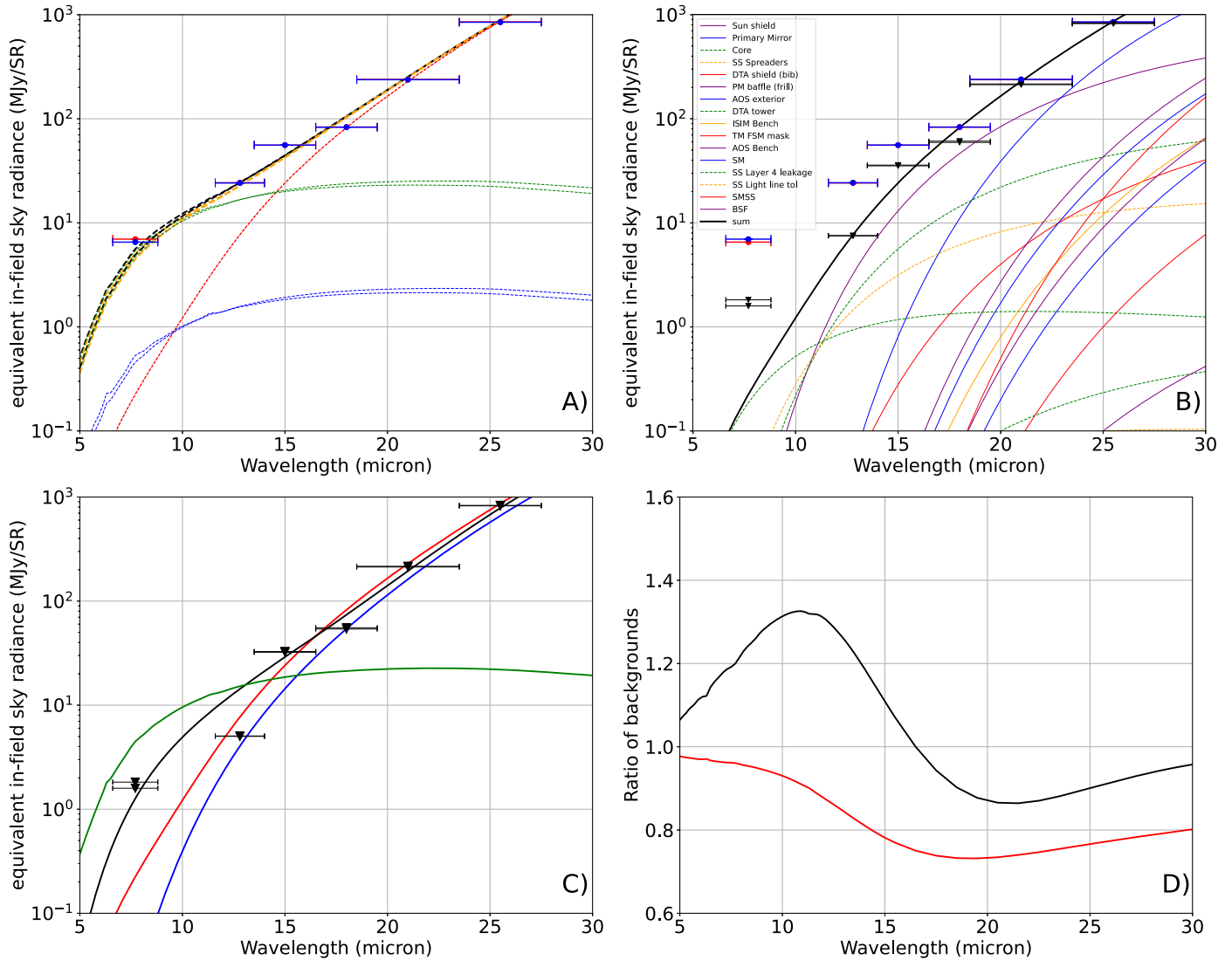

**Figure 7.** The mid-infrared backgrounds for JWST and their expected impact on integration time. Panel (A): Filled circles show the MIRI imaging measurements from the thermal stability test, red points show the hot attitude, and blue points show the cold attitude. Horizontal error bars show each filter bandwidth. The predicted summed in-field zodiacal and ISM backgrounds are shown by the dashed green curves, and the predicted stray light is shown by the dashed blue curves. The predicted total background is shown by the dashed black curve. Panel (B): The same MIRI photometry as in the left panel (blue and red circles), as well as the photometry after subtraction of all the predicted nonthermal components of the background (black triangles). Overplotted is the prelaunch thermal model (black curve) and its subcomponents. The subcomponents are color-coded and are plotted with a solid line if the four-wavelength fit converged. They are plotted with a dashed line otherwise. The legend for the components is sorted in decreasing order of predicted flux density at 20 $\mu$m. Panel (C): The thermal-only component of the background, with the same zodi-and-galaxy-subtracted photometry as in panel (B) (black triangles). Plotted for comparison are prelaunch predictions for the thermal spectrum at the beginning of life (shortly after launch; blue curve), end of life (red curve), and our modification of the end-of-life curve to better fit the commissioning data (black curve). For comparison, the in-field zodiacal component of the background is plotted (green curve) for the 1.2 minimum zodi pointing on the date that field was observed during commissioning. Panel (D): Ratio of the total backgrounds (astrophysical plus thermal) for several scenarios. In the background-limited case, this ratio approximates the fractional change in exposure time to reach the same S/N. To make the black curve, we added the post-commissioning thermal curve from panel (C) to the expected astronomical backgrounds, and divided by the sum of the prelaunch end-of-life prediction and the expected astronomical backgrounds. To make the red curve, we did the same with the predicted prelaunch thermal curves for the beginning-of-life and end-of-life performance. The zodiacal and ISM components used are for the 1.2 minimum zodi field on the day it was observed by JWST during commissioning.

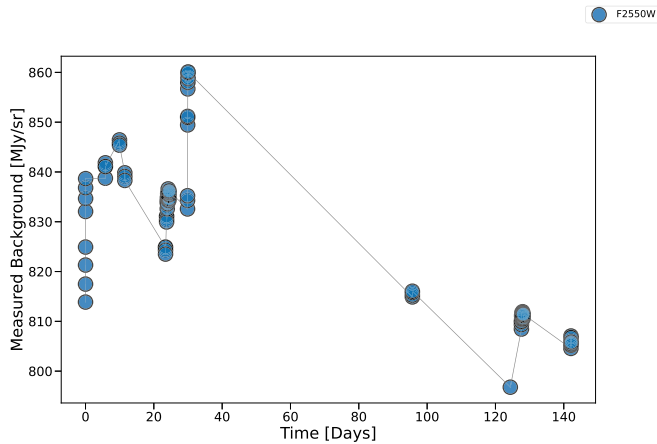

**Figure 8.** JWST background measurements at 25.5 μm, spanning a period of about 145 days, starting in 2022 mid-March. Individual points represent the median across the MIRI imager field of view. The total range of the variation is 7%.

Operationally, both the astrophysical stray-light background and the mid-infrared thermal background are sufficiently close to prelaunch expectations that no modifications are needed to approved Cycle 1 observing programs. The proposal planning software was updated[23] to reflect these on-orbit backgrounds to support the Cycle 2 call for proposals. There is some evidence from the limited number of deep-field observations that the stray-light spectrum has a different shape than the prelaunch expectations, with more of the flux concentrated between 1.5 and 2 μm, and relatively less flux at shorter and longer wavelengths. It is unclear whether this is a real effect or may be due to flux calibration issues that will be resolved later. Passive monitoring of the astrophysical stray-light levels during normal science operations would characterize the stray-light behavior and support further updates to the backgrounds used by the planning tools, as needed.

The thermal background spectrum should also be monitored passively from science observations. There should be a slow, modest degradation in the performance of the sunshield over the mission lifetime due to two causes: (a) accumulated damage of the sunshield by micrometeorites, and (b) damage to the hottest layer of the sunshield due to solar exposure. Gradual degradation should be inferrable through both slowly rising temperatures onboard the observatory (particularly on the four primary mirror segments that are closest to the sunshield), and through a slowly increasing thermal background spectrum. No such degradation has yet been detected; it will be something to watch for in the coming years of science operations. As needed, the thermal background spectrum in the planning tools should be updated.

Taking a broader perspective, the measurements of JWST's near-infrared and mid-infrared backgrounds presented in this paper make clear the success of the multipronged, decades-long efforts to control stray light (Section 3.1) and thermal emission (Section 3.2). Multiple baffles including a frill around the entire primary mirror and rigorous control of contamination have clearly succeeded; the stray-light levels are not only lower than requirements, but lower than expectations. Similarly, given the observed thermal backgrounds we measure, the sunshield and insulation around the core region must have deployed as designed, and other thermal mitigations described in Section 3 must also have succeeded.

Potential lessons learned for future missions are the importance, at the beginning, of setting realistic contamination levels and including integration and test in the contamination budget. JWST has shown that cleaning mirrors can keep contamination within limits and thereby relax their environmental exposure specifications. The success of integrated thermal and optical modeling, validated by a ground test, of a system that cannot be tested as it flies is also clear.

JWST's science requirements were built around being a deep-field machine that would be limited by the irreducible emission from our solar system, not by the telescope itself. This paper demonstrates that this is indeed the case for $\lambda < 12.5$ μm. In moderately deep fields such as the 1.2 minimum zodi benchmark field, at 2–3 μm, from stray-light considerations alone, JWST's integration times can be ∼8% shorter than predicted prelaunch predictions, and up to 22% shorter than if the system just met the stray-light requirements. This result of lower-than-expected stray-light result is just one of several key ways in which JWST exceeds its design requirements, and therefore will be even more powerful than designed.

We are so grateful for the thousands of people around the world who designed, built, commissioned, and operate JWST, and for the support of their families and friends.

This work is based on observations made with the NASA/ESA/CSA James Webb Space Telescope. The data were obtained from the Mikulski Archive for Space Telescopes at the Space Telescope Science Institute, which is operated by the Association of Universities for Research in Astronomy, Inc., under NASA contract NAS 5-03127 for JWST. These observations are associated with programs (PID) 1024, 1027, 1028, 1039, 1040, 1051, 1052, 1052, 1324, 1448, 1521, 1947, 2666, 2736, and 2738.

*Facilities:* JWST(NIRCam, NIRISS, NIRSpec, MIRI).

*Software:* Astropy (Collaboration et al. 2013, 2018, 2022), Numpy (Walt & Colbert 2011), Pandas (McKinney 2010), Matplotlib (Hunter 2007), Jupyter (Kluyver et al. 2016).

# Appendix

This appendix contains Figure 9.

---

[23] Pandeia 2.0, the JWST exposure time caulculator 2.0, and the JWST backgrounds tool 3.0.

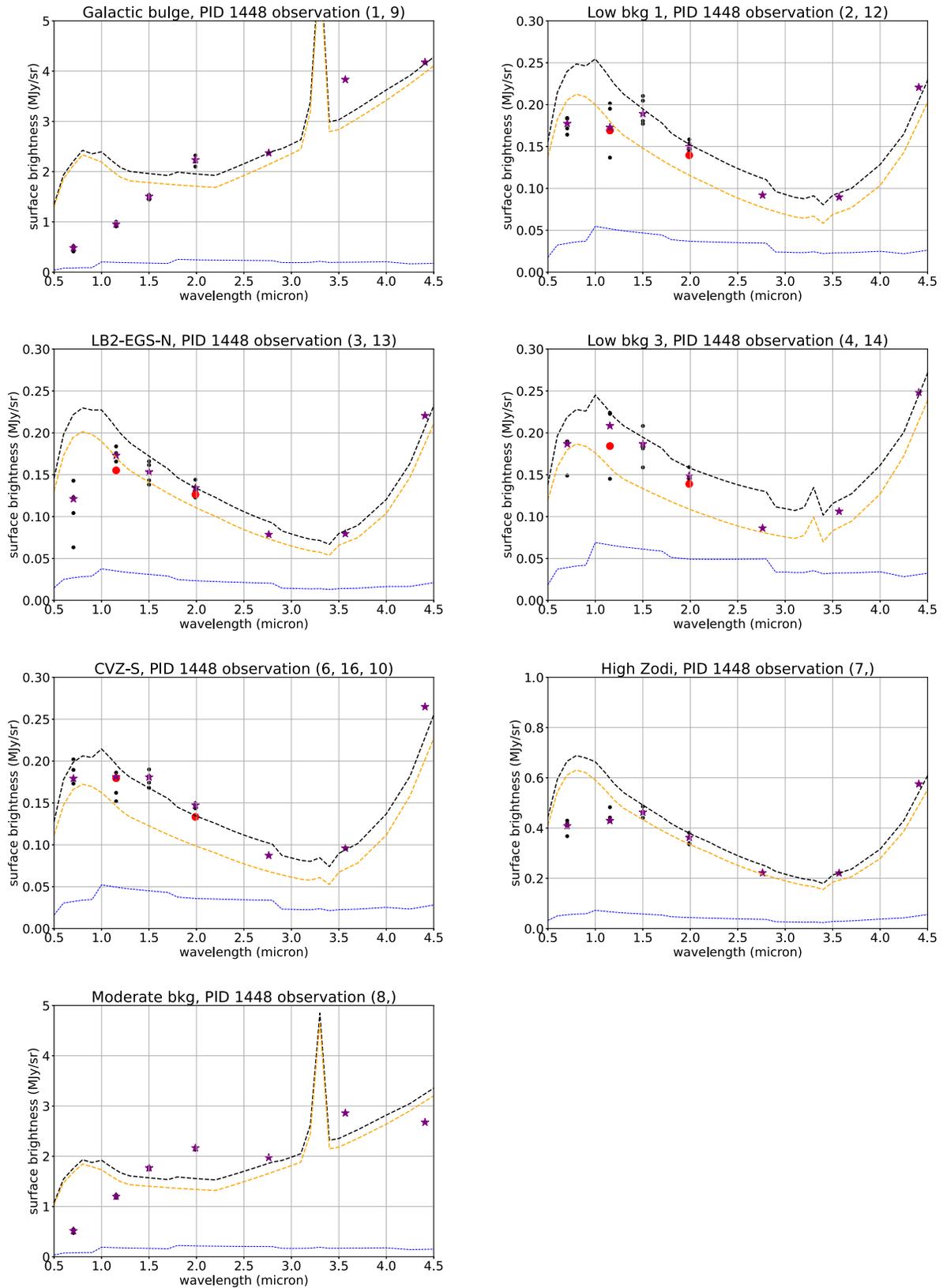

**Figure 9.** Appendix figure: Backgrounds for all the stray-light commissioning pointings listed in Table 2. Symbols are the same as in Figure 4.

## ORCID iDs

Jane R. Rigby https://orcid.org/0000-0002-7627-6551
Paul A. Lightsey https://orcid.org/0000-0001-9185-1393
Macarena García Marín https://orcid.org/0000-0003-4801-0489
Alistair Glasse https://orcid.org/0000-0002-2041-2462
Michael W. McElwain https://orcid.org/0000-0003-0241-8956
George H. Rieke https://orcid.org/0000-0003-2303-6519
Alberto Noriega-Crespo https://orcid.org/0000-0002-6296-8960
Irene Shivaei https://orcid.org/0000-0003-4702-7561
Tea Temim https://orcid.org/0000-0001-7380-3144
Chris J. Willott https://orcid.org/0000-0002-4201-7367